\documentclass[12pt]{article}
\usepackage{graphicx}
\usepackage{xcolor}
\usepackage[colorlinks=true,linkcolor={blue!50!black},citecolor={blue!50!black},urlcolor={blue!80!black}]{hyperref}
\usepackage{bbm}
\usepackage{amssymb}
\usepackage{amscd}
\usepackage{amsmath}
\usepackage{appendix}
\usepackage{appendix}

\begin{document}
\begin{titlepage}


\bigskip \vspace{3\baselineskip}

\begin{center}
{\bf \large 
Observable Zitterbewegung in Curved Spacetimes}

\bigskip

\bigskip

{\bf  Archil Kobakhidze$^{\rm a}$, Adrian Manning$^{\rm a}$ 
and Anca Tureanu$^{\rm b}$   \\}

\smallskip

{ \small \it
$^{\rm a}$ARC Centre of Excellence for Particle Physics at the Terascale, \\
School of Physics, The University of Sydney, NSW 2006, Australia \\ 
$^{\rm b}$Department of Physics, University of Helsinki,
P.O. Box 64, 00014 Helsinki, Finland \\ 
E-mails:  archilk@physics.usyd.edu.au, a.manning@physics.usyd.edu.au, anca.tureanu@helsinki.fi
\\}

\bigskip
 
\bigskip

\bigskip

{\large \bf Abstract}

\end{center}
\noindent {\small
Zitterbewegung, as it was originally described by Schr\"odinger, is an unphysical, non-observable effect. We verify whether the effect can be observed in non-inertial reference frames/curved spacetimes, where the ambiguity in defining particle states results in a mixing of positive and negative frequency modes. We explicitly demonstrate that such a mixing is in fact necessary to obtain the correct classical value for a particle's velocity in a uniformly accelerated reference frame, whereas in cosmological spacetime a particle does indeed exhibit Zitterbewegung.} 

\vspace{2cm}
\end{titlepage}

\section{Introduction} 
Zitterbewegung (``trembling motion") of a free relativistic electron described by the Dirac equation was theorised by Schr\"odinger in 1930 \cite{zitter}. It has been subsequently understood that Zitterbewegung is actually an unphysical, non-observable effect. In one-particle (quantum mechanical) interpretation one can demonstrate this by performing specific unitary transformations suggested by Foldy and Wouthuysen \cite{Foldy:1949wa}. The effect disappears in the Foldy-Wouthuysen representation, and this suggests that a naive superposition of negative and positive energy solutions in the original Dirac representation is erroneous. A full resolution comes from the consistent second-quantized treatment according to which any real physical particle (antiparticle) carries only positive energy and thus no superposition of negative and positive energy states is possible. The effect of virtual particles is taken into account to describe the physical electron (mass, charge and wave function renormalizations) and, obviously, does not affect its free 
motion\footnote{Despite this clear-cut picture some researchers still advocate Zitterbewegung as a real effect and some even claim its experimental discovery \cite{zitterexp}. In our opinion these claims are erroneous (see also the criticism in \cite{antizitter}). }. 

In this paper we demonstrate that Zitterbewegung may show up as a physical effect in certain reference frames which cannot be defined globally on an  entire spacetime. In such spacetimes, the definition of particle states become observer-dependent and thus ambiguous. Well-known examples  are reference frames related to stationary observers in black hole spacetimes and uniformly accelerating (Rindler) observers in flat Minkowski spacetime. Such observers perceive the Minkowski vacuum state as  being a thermal radiation coming from the event horizon \cite{Hawking:1974rv, Fulling:1972md}, which is due to the non-trivial mixing of positive and negative frequency modes of quantized fields. Similar mixing and the  related particle production occurs in time-dependent (cosmological) spacetimes \cite{Parker:1969au}. We demonstrate that precisely this mixing of negative and positive frequency modes causes Zitterbewegung of a  Dirac fermion in cosmological spacetime, whereas in Rindler spacetime it leads to a one-particle expectation value for velocity which exactly reproduces its classical counterpart.  In what follows we restrict our discussion to (1+1)-dimensional toy models, since  this case is technically simpler and captures the essential physics behind the phenomenon. We consider other physically interesting spaces elsewhere.

The paper is organised as follows. In the next section we set up our notations and show that no Zitterbewegung appears in inertial reference frames. Section 3 is devoted to the calculation of one-particle velocity expectation value for uniformly accelerated observer. Finally, in Section 4 we demonstrate explicitly the physical Zitterbewegung within a toy cosmological model. The last section is devoted to conclusions and some useful formulae are collected in the Appendix.  
 
\section{No Zitterbewegung for inertial observers}
For the sake of completeness and to set up our notations we start our discussion by demonstrating the absence of Zitterbewegung in inertial reference frames. (1+1)-dimensional Minkowski spacetime is parameterized by Cartesian coordinates $(t,x)$ and the interval reads:
\begin{equation}
ds^2=dt^2-dx^2~.
\label{1}
\end{equation}
The Dirac spinor in (1+1) dimensions\footnote{For a detailed presentation of Clifford algebras and spinors in arbitrary spacetime dimensions, see, for example, Ref. \cite{MRT}.} is a two-component complex field $\psi$, which satisfies the Dirac equation
\begin{equation}
(i\gamma^\mu\partial_\mu -m) \psi= 0,
\label{2}
\end{equation}
where the $(1+1)$-dimensional $\gamma$-matrices satisfy the anticommutation relation
$$
\{\gamma^\mu,\gamma^\nu\}=2g^{\mu\nu}\mathbf{1}_{2\times 2}.
$$ For this calculation, we will use the following representation for the $\gamma$-matrices:
\begin{eqnarray}
\gamma^{0}=\left(
\begin{tabular}{cc}
0 & 1 \\
1 & 0
\end{tabular}
\right)~, ~~
\gamma^{1}=\left(
\begin{tabular}{cc}
0 & 1 \\
-1 & 0
\end{tabular}
\right)~.
\label{3}
\end{eqnarray}
The Dirac equation is obtained as Euler--Lagrange equation from the Lagrangian:
\begin{equation}\label{Lagrangian}
{\cal L}=i\bar\psi\gamma^\mu\partial_\mu\psi-m\bar\psi\psi.
\end{equation}
The canonical formalism leads to the Hamiltonian density
\begin{equation}\label{Hamiltonian}
{\cal H}=\pi\partial_0\psi-{\cal L}=\bar\psi(-i\gamma^1\partial_1+m)\psi=\bar\psi\gamma_0\psi,
\end{equation}
where $\pi(x)=\frac{\partial{\cal L}}{\partial(\partial_0\psi)}$ and in the last equality was used the equation of motion \eqref{2}. The same Hamiltonian is of course obtained from Noether's theorem, in view of the translational invariance of the Lagrangian \eqref{Lagrangian}, which leads to the conserved energy-momentum tensor
\begin{equation}
T^{\mu\nu}=i\bar\psi\gamma^\mu\partial^\nu\psi-g^{\mu\nu}{\cal L},
\end{equation}
whose component $T^{00}$ is the Hamiltonian density \eqref{Hamiltonian}.

By solving Dirac's equation \eqref{2} (see, for example, Ref. \cite{BD}), one obtains the normalized positive/negative frequency $\left(\omega_p=\sqrt{p^2+m^2}\right)$ solutions, $\psi^{(\pm)}$, which are the familiar plane waves: 
\begin{eqnarray}
\psi^{(\pm)}_p=\frac{1}{\sqrt{2\omega_p}}\left(
\begin{tabular}{c}
$\sqrt{\omega_p\mp p}$ \\
$\pm\sqrt{\omega_p\pm p}$
\end{tabular}\right){\rm e}^{\mp i\omega_pt+ipx}~.
\label{4}
\end{eqnarray}
Thus, a free quantum field as described by an inertial observer in Minkowski coordinates reads:
 \begin{equation}
 \hat \psi_{\rm M}(t,x)=\int dp \frac{1}{\sqrt{2\pi}} \left(\hat a(p)\psi^{(+)}_p+\hat b^{\dagger}(-p)\psi^{(-)}_p\right)~, 
 \label{5}
 \end{equation}
where the annihilation (creation) operators $\hat a (\hat a^{\dagger})$, $\hat b (\hat b^{\dagger})$, respectively for particles and antiparticles, satisfy the standard anticommutation relations:
\begin{equation}
\lbrace \hat a(p), \hat a^{\dagger}(p')\rbrace=\delta(p-p')~,~~   \lbrace \hat b(p), \hat b^{\dagger}(p')\rbrace=\delta(p-p')~,
\label{6}
\end{equation}
 other anticommutators being zero. Particle states are defined as excitations over the vacuum state $\vert0_{\rm M}\rangle$:
 \begin{equation}
 \hat a(p)\vert0_{\rm M}\rangle=\hat b(p)\vert0_{\rm M}\rangle=0~,~~\forall  p~.
 \label{7}
 \end{equation} 
 In particular, particle and antiparticle with momentum $p$ are described respectively by the states, 
 \begin{equation}
 \vert  p\rangle = \hat a^{\dagger}(p)\vert0_{\rm M}\rangle~,~~\tilde{\vert  p\rangle} = \hat b^{\dagger}(p)\vert0_{\rm M}\rangle~, 
 \label{8}
 \end{equation}
which are orthogonal to each other, $\langle p \tilde{\vert  p\rangle}=0$, and 
both carry positive energy $E=+\sqrt{p^2+m^2}$. 

Next, we define a particle's velocity as follows. The Dirac Lagrangian \eqref{2} is invariant under $U(1)$ global transformations, which leads by Noether's theorem to the conservation of the current
$j^\mu=\bar\psi\gamma^\mu\psi$. Upon quantization, the current becomes an operator whose components are the 
current and charge density operators:
\begin{eqnarray}
\hat j^1&= &:\hat \psi^{\dagger}\gamma^0\gamma^1\hat \psi:~,~~ \\ 
\hat j^0&=& :\hat \psi^{\dagger}\hat \psi:~,~~
\label{10}
\end{eqnarray}
where $:...:$ denotes normal ordering of operators and $\mu = \{0,1\}$ are the Lorentz vector indices in (1+1)-dimensional spacetime.  Expressing the operators in terms of creation and annihilation operator, it is then easy to show that the velocity of a particle with momentum $p$ and energy $\omega_p$ is:
\begin{equation}
v \equiv \frac{\langle p\vert \hat j^1\vert p\rangle}{\langle p\vert \hat j^0\vert p\rangle} = \frac{p}{\omega_p}~.
\label{11}
 \end{equation}
This  expression is just the classical velocity of a relativistic particle.  Hence,  no Zitterbewegung term appears within the consistent description of the Dirac particle in inertial reference frames.  

\section{No Zitterbewegung in an uniformly accelerated reference frame}
In this section we consider Dirac fermions in a uniformly accelerated reference frame.  An observer moving with a constant positive acceleration, $a$, along the $x$-direction in (1+1)-dimensional flat Minkowski spacetime, accounts for an event horizon. Such a non-inertial frame is described by the  Rindler spacetime, with the following line element: 
\begin{equation}
ds^2=a^2\chi^2 d\tau^2 - d\chi^2~.
\label{14}
\end{equation}
For details about quantum field theory in Rindler space, see, for example, Ref. \cite{Birrell:1982ix}. The Rindler coordinates $(\tau, \chi)$ can be written in terms of Cartesian coordinates $(t,x)$, which parametrize the flat Minkowski spacetime, as follows:
\begin{equation}
\tau=\frac{1}{a}{\rm arctanh}\left(\frac{t}{x} \right)~,~~\chi=\sqrt{x^2-t^2}~.
\label{15}
\end{equation}  
It is obvious from the above equations that the Rindler coordinates cover only the $|x|>t$ quadrant of the full Minkowski spacetime and $\chi \in [0, +\infty]$.  
 
The Dirac equation in Rindler spacetime reads:
\begin{equation}
 i \gamma^0 \partial_{\tau}= - i a \chi \gamma^0\gamma^1 \partial_\chi\psi - \left(\frac{ia}{2} \gamma^0\gamma^1 - a \chi m\gamma^0 \right) \psi.
\label{16}
\end{equation}
The positive/negative frequency solutions to the above equation can be found in terms of modified Bessel function of the second kind:
\begin{equation}
\psi_{\Omega}^{(\pm)}=\sqrt{\frac{m \cosh (\frac{\pi}{a}\Omega)}{\pi^2a}} e^{\mp i \Omega \tau} \begin{pmatrix}	 K_{\pm i \frac{\Omega}{a} + \frac{1}{2}} (m \chi) \\ i K_{\pm i\frac{\Omega}{a} - \frac{1}{2}}(m \chi) \end{pmatrix}~, 
\label{17}
\end{equation}
where we have adopted the normalisation condition: $\int_0^{\infty}\left(\psi_{\Omega'}^{(\pm)}\right)^{\dagger}\psi_{\Omega}^{(\pm)}d\chi=\delta\left(\Omega'-\Omega\right)$ and $\Omega \geq 0$. The mathematical details of these calculations can be found in the Appendix. 

Given the mode solutions (\ref{17}) we construct the Rindler quantum field as: 
\begin{equation}
\hat \psi_{\rm R}(\tau, \chi)=\int_0^{\infty}d\Omega\left(\hat A_{\Omega}\psi^{(+)}_{\Omega}+\hat B^{\dagger}_{\Omega}\psi^{(-)}_{\Omega}\right)~,
\label{18}
\end{equation}
where $\hat A (\hat B)$ and $\hat A^{\dagger}(\hat B^{\dagger})$ are particle (antiparticle) creation and annihilation operators, respectively, acting on the Fock space as defined by the Rindler observer, i.e., $\hat A_{\Omega}\vert 0_{\rm R}\rangle=\hat B_{\Omega} \vert 0_{\rm R}\rangle=0,~ \forall \Omega$,  $\vert 0_{\rm R}\rangle$ being the Rindler vacuum state, etc. We then define the Rindler current and charge density operators as:
\begin{eqnarray}
\hat j^1&= & a\chi :\hat \psi_{\rm R}^{\dagger}\gamma^0\gamma^1\hat \psi_{\rm R}:~,~~  
\label{19} \\
\hat j^0&=& :\hat \psi_{\rm R}^{\dagger}\hat \psi_{\rm R}:~,~~
\label{20}
\end{eqnarray} 
where  the normal ordering is defined with respect to the Rindler operators. Note also that the time evaluation is assumed with respect to the Rindler coordinate time $\tau$, which is related to the proper time $\tau_p$ as $a\chi d\tau = d\tau_p$. 

According to (\ref{15}), the uniformly accelerated frame can be viewed as a Lorentz boosted inertial reference frame with an instantaneous velocity $a\tau$. Hence we can relate the boosted Minkowski modes to the Rindler modes via Bogoliubov transformations, i.e.:
\begin{eqnarray}
{\rm e}^{-\frac{1}{2} \gamma^0 \gamma^1 a \tau}  \psi_{p}^{(+)}(t(\tau,\chi), x(\tau,\chi)) = \int_{0}^{\infty} d\Omega \; \alpha_{\Omega,p} \, \psi_\Omega^{(+)} + \beta_{\Omega,p} \, \psi_\Omega^{(-)}~, 
\label{21} \\
{\rm e}^{-\frac{1}{2} \gamma^0 \gamma^1 a \tau}  \psi_{p}^{(-)}(t(\tau,\chi), x(\tau,\chi)) = \int_{0}^{\infty} d\Omega \; \alpha^*_{\Omega,p} \, \psi_\Omega^{(-)} + \beta^*_{\Omega,p} \, \psi_\Omega^{(+)}~.
\label{22}
\end{eqnarray}
Using these equations and the relations given in the Appendix we have computed the Bogoliubov coefficients (see the details in the Appendix):
 \begin{equation}
 \alpha_{\Omega, \theta_p} = \frac{1-i}{2} \frac{e^{\frac{i \Omega \theta_p}{a}} e^{\frac{\pi \Omega}{2a}}}{\sqrt{2a\pi m \cosh(\theta_p) \cosh(\frac{\Omega}{a} \pi) }}~,~~ \beta_{\Omega,\theta_p} = \alpha_{-\Omega,\theta_p}~. 
 \label{23}
 \end{equation}
 In (\ref{23}) we have parameterized the Minkowski frequency and momentum as: 
\begin{equation}
\omega_p = m \cosh(\theta_p)~,~~p = m \sinh(\theta_p)~.
\label{24} 
\end{equation} 		
 Given the above Bogoliubov coefficients we can map Rindler ($\hat A$,$\hat B$) and Minkowski ($\hat a$, $\hat b$) operators:  
 \begin{eqnarray}
	\hat{A}_\Omega &=   \int_{-\infty}^{\infty} dk \, (\alpha_{\Omega,k} \hat{a}_k + \beta^*_{\Omega,k} \hat{b}_k^\dagger)~, 
\label{25}	
	\\
\nonumber \\	
	\hat{B}^\dagger_\Omega &=  \int_{-\infty}^{\infty} dk \, (\beta_{\Omega,k} \hat{a}_k + \alpha^*_{\Omega,k} \hat{b}_k^\dagger)~.  
\label{26}
 \end{eqnarray}
 As described in the Appendix, we finally obtain the expectation values for the Rindler current and charge density operators given generically by (\ref{A31}) (see also for the specific case of the current density (\ref{A32})), respectively, for a one-particle Minkowski state $\vert p \rangle$:
\begin{eqnarray}
\langle p \vert\hat j^1_{\rm R}\vert p \rangle=\int_{-\infty}^{\infty} \int_{-\infty}^{\infty} d\Omega \, d\Omega' {\psi^{(+)}_\Omega}^\dagger \gamma^0 \gamma^1 a \chi \, \psi^{(+)}_{\Omega'} \; \alpha_{\Omega,p}^* \alpha_{\Omega',p} = \frac{a\chi \sinh(\theta_p - a\tau)}{\pi \cosh(\theta_p)}~, 
\label{27} \\
\nonumber \\
\langle p \vert\hat j^0_{\rm R}\vert p \rangle= \int_{-\infty}^{\infty} \int_{-\infty}^{\infty} d\Omega \, d\Omega' {\psi^{(+)}_\Omega}^\dagger O \, \psi^{(+)}_{\Omega'} \; \alpha_{\Omega,p}^* \alpha_{\Omega',p}=\frac{\cosh(\theta_p - a\tau)}{\pi \cosh(\theta_p)}~.
\label{28}
\end{eqnarray} 
Importantly, the limit of integration in the above equations reflect the fact that both the positive and negative frequency solutions do contribute to the final result. Also we note that the computed expectation values are finite due to the Rindler normal ordering adopted in (\ref{19}) and (\ref{20}). With these in hand, we compute the velocity of a particle in a uniformly accelerated reference frame, 
\begin{equation}
v_{\rm R}=\left. \frac{\langle p \vert\hat j^1_{\rm R}\vert p \rangle}{\langle p \vert\hat j^0_{\rm R}\vert p \rangle}\right \vert_{a\chi=1}=  \tanh(\theta_p - a\tau)~, 
\label{29}
\end{equation}      		
 which is precisely the classical velocity. Hence, we see that, similar to inertial frames, there is no Zitterbewegung effect in uniformly accelerated reference frames.   
 
 \section{Cosmological Zitterbewegung} 
Let us consider now a toy (1+1)-dimensional  cosmological model (see. e.g., Ref. \cite{Birrell:1982ix}) with the line element 
\begin{equation}
ds^2 = a^2(\eta) (d\eta^2 - dx^2), 
\label{30}
\end{equation}
which describes a change of the scale factor from a constant, $a_0 - \chi$, in the infinite past ($\eta \rightarrow -\infty$) to another constant $a_0 + \chi$ in the infinite future ($\eta \rightarrow  \infty$):
\begin{equation}
a(\eta) = a_0 + \chi \tanh(\varepsilon \eta) .
\end{equation}
This model describes an asymptotically static spacetime which undergoes a period of smooth expansion.
The Dirac equation in this spacetime now reads
\begin{equation}
	\left[	i\gamma^\mu \partial_\mu + i \frac{1}{2} \frac{a'(\eta)}{a(\eta)} \gamma^0 -ma(\eta) \right] \psi = 0 ~,
	\label{eq:cosmoeom}
\end{equation}
where the prime denotes the derivative with respect to $\eta$. 
In the following calculations we shall use the notations:
\begin{equation}
        \begin{aligned}
		a_{in/out} &= (a_0 \mp \chi), \\
		\omega_{in/out} &= \sqrt{p^2 + m^2a_{in/out}^2}, \\
		\omega_{\pm} &= \frac{1}{2}(\omega_{out} \pm \omega_{in}).
        \end{aligned}
\end{equation}
The solutions to this equation can be found using the Dirac representation for the $\gamma$-matrices:
\begin{eqnarray}
\gamma^{0}=\left(
\begin{tabular}{cc}
1 & 0 \\
0 &-1 
\end{tabular}
\right)~, ~~
\gamma^{1}=\left(
\begin{tabular}{cc}
0 & 1 \\
-1 & 0
\end{tabular}
\right)~.
\end{eqnarray}
The positive energy solutions which match the flat-space plane waves in the infinite past, denoted below as ``in" states, are found to be (see Appendix for a brief discussion):
\begin{equation}
        \psi_{in}^+ =  \; (e^{\varepsilon \eta} + e^{-\varepsilon \eta })  e^{- i \omega_+ \eta} e^{-\left(1 + \frac{i \omega_-}{\varepsilon}\right) \ln (2 \cosh (\varepsilon \eta))} \frac{a(\eta)^{-\frac{1}{2}}}{\sqrt{2 \omega_{in}}} \begin{pmatrix} U_1 \\ U_2 \end{pmatrix} e^{ipx} ~,
	\label{eq:instate}
\end{equation}
with
\begin{equation}
        \begin{aligned}
                U_1 = & \sqrt{\omega_{in} + ma_{in}} \;  F_1,  \\
                U_2 = & \frac{|p|}{p} (\omega_{in} - ma_{in})^{-\frac{1}{2}}  \\
                & \times \;\Big[ F_1 \left[ - ma_0 + \tanh(\varepsilon \eta) (i \varepsilon - m \chi) + \frac{i}{2}\left[ (i \omega_{in} - \varepsilon)(\tanh(\varepsilon \eta) -1) - (i \omega_{out} +                \varepsilon)(\tanh(\varepsilon \eta) + 1) \right] \right] \\
                & - \frac{F_2}{i \omega_{in} - \varepsilon} \left[ 4i\left[ (\omega_-)^2 - m^2 \chi^2      \right] + 4 \varepsilon(\omega_- - m\chi) \right] \left( \tanh(\varepsilon \eta) +1 \right) \left(      \tanh(\varepsilon \eta) -1  \right) \Big] ~.
                \end{aligned}
\end{equation}
Here we have used the following condensed notation:
\begin{equation}
        \begin{aligned}
                F_1 &= {}_2F_1 \left(\frac{i}{\varepsilon} \left(\omega_- - m \chi \right) ,  1 + \frac{i}{\varepsilon} (\omega_- + m\chi); 1 - \frac{i \omega_{in}}{\varepsilon};          \frac{1}{2} \tanh(\varepsilon \eta) + \frac{1}{2} \right), \\
                F_2 &= {}_2F_1 \left(2 + \frac{i}{\varepsilon} \left( \omega_- + m \chi \right) , 1 + \frac{i}{\varepsilon} (\omega_- - m\chi); 2 - \frac{i \omega_{in}}{\varepsilon}; \frac{1}{2} \tanh(\varepsilon \eta) + \frac{1}{2} \right) ,
        \end{aligned}
\end{equation}
with ${}_2F_1(a,b;c;z)$ being the ordinary hypergeometric function.
The states which solve the Dirac equation and match the plane wave solutions in the infinite future are denoted as ``out" states and the positive energy solutions are found to be:
\begin{equation}
        \psi_{out}^+ =  \; (e^{\varepsilon \eta} + e^{-\varepsilon \eta })  e^{- i \omega_+ \eta} e^{-    (1 + \frac{i \omega_-}{\varepsilon}) \ln (2 \cosh (\varepsilon \eta))} \frac{a(\eta)^{-\frac{1}{2}}}{\sqrt{2 \omega_{out}}} \begin{pmatrix} U_1 \\ U_2 \end{pmatrix} e^{i px} ~,
	\label{eq:outstate}
\end{equation}
with
\begin{equation}
        \begin{aligned}
                U_1 = & \sqrt{\omega_{out} + ma_{out}} \;  \tilde    F_1 \\
                U_2 = & \frac{|p|}{p} (\omega_{out} - ma_{out})^{-\frac{1}{2}}  \\
                & \times \;\Big[ \tilde F_1 \left[ - ma_0 + \tanh(\varepsilon \eta) (i \varepsilon - m  \chi) + \frac{i}{2}\left[ (i \omega_{in} - \varepsilon)(\tanh(\varepsilon \eta) -1) - (i \omega_{out} +        \varepsilon)(\tanh(\varepsilon \eta) + 1) \right] \right] \\
                & - \frac{ \tilde F_2}{i \omega_{in} - \varepsilon} \left[ 4i\left[ (\omega_-)^2 - m^2     \chi^2 \right] + 4 \varepsilon(\omega_- - m\chi) \right] \left( \tanh(\varepsilon \eta) +1 \right)      \left( \tanh(\varepsilon \eta) -1  \right) \Big] .
                \end{aligned}
\end{equation}
Here we have introduced the following notation
\begin{equation}
	\begin{aligned}
		\tilde F_1 &= 	{}_2F_1\left[   \frac{i}{\varepsilon} (\omega_- - m \chi) , 1 + \frac{i}{\varepsilon} (\omega_- + m\chi) ; 1 + \frac{i \omega_{out}}{\varepsilon}; \frac{1}{2} - \frac{1}{2} \tanh (\varepsilon \eta) \right] \\
    \tilde F_2 &= {}_2F_1 \left[ 2 + \frac{i}{ \varepsilon} \left( \omega_- + m \chi \right) , 1 + \frac{i}{\varepsilon} (\omega_- - m\chi); 2 + \frac{i \omega_{out}}{\varepsilon}; \frac{1}{2} -  \frac{1}{2}\tanh(\varepsilon \eta)\right]
	\end{aligned}
\end{equation}
The negative energy states can be found by using the charge conjugation operator as follows
\begin{equation}
	\psi^- = C{\psi^*}^+ ~.
\end{equation}
In our 2-D Dirac representation, the charge conjugation operator is
\begin{equation}
C=\left(
\begin{tabular}{cc}
0 & 1 \\
1 & 0
\end{tabular}
\right)~.
\end{equation}
The field operator can be expanded in terms of the above mode functions as follows:
\begin{eqnarray}
 \hat \psi_{in}=\int dp \frac{1}{\sqrt{2\pi}} \left(\hat a(p)\psi^{(+)}_{p;in}+\hat b^{\dagger}(-p)\psi^{(-)}_{p;in}\right)~,& {\rm for}~ \eta<\eta_0~, \\ 
\label{34} \nonumber \\
\hat \psi_{out}=\int dp \frac{1}{\sqrt{2\pi}} \left(\hat A(p)\psi^{(+)}_{p;out}+\hat B^{\dagger}(-p)\psi^{(-)}_{p; out}\right)~,&{\rm for}~ \eta>\eta_0~. 
\label{35}
\end{eqnarray}
The two sets of creation and annihilation operators in the above equations are related to each other via the Bogoliubov transformations:
\begin{eqnarray}
	\hat{A}(p) = \alpha(p)\hat{a}(p) + \beta^*(-p) \hat{b}^\dagger(-p)~, \label{36}  \\
	\nonumber \\
	\hat{B}^\dagger (p) = \beta(-p) \hat{a}(-p) + \alpha^*(p) \hat{b}^\dagger (p)~,
	\label{37}
\end{eqnarray}
where 
 \begin{eqnarray}
\alpha(p)  = \frac{\sqrt{\omega_{out}}\sqrt{\omega_{in} + ma_i} }{\sqrt{\omega_{in}} \sqrt{\omega_{out} + m a_{out}}} \frac{\Gamma(1 - \frac{i \omega_{in}}{\varepsilon}) \Gamma(-\frac{i \omega_{out}}{\varepsilon})}{\Gamma(1 - \frac{i}{\varepsilon}(\omega_{+} - m\chi) \Gamma(-\frac{i}{\varepsilon}(\omega_{+} + m\chi))} ~, \label{eq:cosmoalpha} \\ 
\nonumber \\
		\beta(p)  =-\frac{|p|}{p} \frac{\sqrt{\omega_{out}} \sqrt{\omega_{in} + ma_i}}{\sqrt{\omega_{in}}\sqrt{\omega_{out}-ma_{out}}} \frac{\Gamma(1 - \frac{i \omega_{in}}{\varepsilon}) \Gamma(\frac{i \omega_{out}}{\varepsilon})}{\Gamma(\frac{i}{\varepsilon}(\omega_{-} - m\chi) \Gamma(1 +\frac{i}{\varepsilon}(\omega_{-}  + m\chi))} ~, 
	\label{eq:cosmobeta}
\end{eqnarray} 

Assume now that we prepare a one-particle (Heisenberg) state with momentum $p$ at some earlier time $\eta \rightarrow -\infty$: $\vert p \rangle =a^{\dagger}(p)\vert 0_{M;in}\rangle $, and we wish to compute the velocity of this state at some later time $\eta \rightarrow \infty$, that is $$v={\langle p\vert\hat j^1_{out}\vert p\rangle\over\langle p\vert\hat j^0_{out}\vert p\rangle},$$ where
\begin{equation}
\hat j^{\mu}_{out}=:\hat \psi_{out}^{\dagger}\gamma^0\gamma^{\mu}\hat \psi_{out} :
\label{40}
\end{equation}
The velocity can be calculated generically, and is found to be:
\begin{eqnarray}
	v=&\frac{|p|}{\omega_{out}} \left[ \frac{1}{\tanh(\frac{\pi}{\varepsilon} \omega_{out}) \tanh(\frac{\pi}{\varepsilon} \omega_{in})}  - \frac{\cosh(\frac{2\pi}{\varepsilon} m \chi)}{\sinh(\frac{\pi}{\varepsilon} \omega_{out}) \sinh(\frac{\pi}{\varepsilon} \omega_{in})}+ \xi + \xi^*\right] ,
	\end{eqnarray}
where we have introduced
\begin{equation}
	\begin{aligned}
	\xi =& e^{2i\omega_{out} \eta} \frac{-\omega_{out} m^2 (a_0 + \chi) \chi}{\pi^3 \varepsilon^3 \sinh(\frac{\pi \omega_{in}}{\varepsilon})} \\
\times & \sinh\left(\frac{\pi}{\varepsilon}(\omega_+ + m \chi)\right)\sinh\left(\frac{\pi}{\varepsilon}(\omega_+ - m \chi)\right)\sinh\left(\frac{\pi}{\varepsilon}(\omega_- + m \chi)\right)\sinh\left(\frac{\pi}{\varepsilon}(\omega_- - m \chi)\right) \\
	\times & \Gamma\left(\frac{i \omega_{out}}{\varepsilon}\right)^2 \Gamma\left(-\frac{i}{\varepsilon}(\omega_+ - m\chi)\right) \Gamma\left(-\frac{i}{\varepsilon}(\omega_+ + m\chi)\right)\Gamma\left(-\frac{i}{\varepsilon}(\omega_- + m\chi)\right)\Gamma\left(-\frac{i}{\varepsilon}(\omega_- - m\chi)\right) .
\end{aligned}
\end{equation}
We can consider the case where the spacetime grows rapidly at one instant in time. This case can be evaluated by taking the limit as $\varepsilon \rightarrow \infty$, where the spacetime approaches a step function and the in-states are instantaneously converted to out-states at time $\eta =0$. In this limit, the velocity becomes:   
\begin{eqnarray}
v=\frac{|p|}{\omega_{out}} \frac{a_{in}a_{out}}{\omega_{in} \omega_{out}} 
\left[\frac{p^2}{a_{in}a_{out}}+m^2 \left[1+ \left(\frac{a_{out}}{a_{in}} - 1\right) \cos\left (2 \omega_{out} \eta \right)\right]\right ]~.
\label{41}
\end{eqnarray}
As envisaged earlier in \cite{Girdhar:2013nsa}, we explicitly encounter  in (\ref{41}) the time-dependent oscillatory term, which is very similar to Schr\"odinger's Zitterbewegung term. The effect vanishes for massless particles as in the standard case. As a consistency check, we note that for $a_{in}=a_{out}$, the Zitterbewegung term disappears from (\ref{41}) and one recovers the usual velocity formula given in (\ref{11}).  

We can further consider the case when the Compton wavelength of the particle is much larger than the width of the spacetime change i.e $\frac{m}{\varepsilon} \ll 1$. In this limit, maintaining only linear terms in $\frac{m}{\varepsilon}$, we find that the Bogoliubov coefficients become
\begin{equation}
	\begin{aligned}
		\alpha_p &= 1 + \mathcal{O}\left(\left(\frac{m}{\varepsilon}\right)^2\right), \\
		\beta_p &= \frac{m}{\varepsilon} \frac{|p|}{p} \frac{\pi \chi}{\sinh(\pi \frac{p}{\varepsilon})} + \mathcal{O}\left(\left(\frac{m}{\varepsilon}\right)^2\right) ~,
	\end{aligned}
	\label{}
\end{equation}
and hence for the velocity expectation value (keeping quadratic $\frac{m}{\varepsilon}$ terms for clarity), we achieve
\begin{equation}
	v = \frac{|p|}{\omega_{out}} \left( 1 + \left(\frac{m}{\varepsilon}\right)^2  \frac{\pi}{2 \sinh(\frac{\pi p}{\varepsilon})} \left[ \frac{(2 a_{out} a_{in} - a_{out}^2 - a_{in}^2)}{2 \sinh(\frac{\pi p}{\varepsilon})} + \frac{a_{out}(a_{out}- a_{in})\varepsilon}{p} \cos(2 \omega_{out} \eta) \right] \right) ~.
\end{equation}
In this small mass limit, it is clear that the classical velocity $\frac{p}{\omega_{out}}$ is achieved and only oscillatory terms of order $(\frac{m}{\varepsilon})^2$ contribute.

\section{Conclusion and outlook}

In this paper we have discussed the Zitterbewegung phenomenon in non-inertial reference frames /curved spacetimes, where the 	positive and negative frequency modes of a quantum field get mixed. We have explicitly demonstrated that this mixing is essential to obtain the classical velocity in the Rindler spacetime, that is, similar to an inertial observer, Zitterbewegung is an unphysical and thus unobservable effect for an uniformly accelerated observer. Within a toy cosmological model, however, we have found that Zitterbewegung can indeed occur.

It certainly is interesting to study other non-inertial reference frames/curved spacetimes in which Zitterbewegung manifests  itself as an observable phenomenon. Besides potential applications to cosmology, this study may offer an intriguing possibility to verify Zitterbewegung in laboratory experiments that explore  analogue curved spacetimes (see, e.g., \cite{Barcelo:2005fc} for a recent review). The obvious challenge for such experiments is to maintain a low enough temperature, so that thermal fluctuations do not overshadow the quantum Zitterbewegung effect. In this regard, trapped ions may constitute a promising experimental set-up for the observation of the cosmological Zitterbewegung effect, providing rapid and controlled expansion/contraction of the ion trap can be achieved \cite{Schutzhold:2007mx}.               

\paragraph{Acknowledgements} We are grateful to Masud Chaichian for useful discussion. The work of AK and AM was partially supported by the Australian Research Council. AK was also supported in part by the Rustaveli National Science Foundation under the projects No. DI/8/6-100/12 and No. DI/12/6-200/13.  The support of the Academy of Finland under the Project no. 136539 is gratefully acknowledged.  

\appendix
\numberwithin{equation}{subsection}
\section{Appendix}
\label{Appendix}
In this Appendix we collect useful formulas and relations used for obtaining the results discussed in the main text.  

\subsection{Normalisation of Rindler modes}
The solutions to the Dirac equation in Rindler spacetime (\ref{16}) are:
\begin{equation}
	\psi^{(\pm)} = N_\Omega e^{\mp i \Omega \tau} \begin{pmatrix} K_{\pm i \frac{\Omega}{a} + \frac{1}{2}} (m\chi) \\ i K_{\pm i \frac{\Omega}{a} - \frac{1}{2}} (m\chi) \end{pmatrix}.
	\label{A11}
\end{equation}
We calculate the normalisation factor, $N_\Omega$, by imposing the normalisation condition:
\begin{equation}
	\int_{0}^{\infty} \left({\psi_{_{\Omega}}^{(\pm)}}\right)^\dagger \, \psi_{_{\Omega'}}^{(\pm)} \, d\chi = \delta (\Omega - \Omega').
	\label{A12}
\end{equation}
To evaluate the normalisation factor, we solve the following integral:
\begin{equation}
	\int_{0}^{\infty} dz \, K_{i\frac{\Omega'}{a} + \frac{1}{2}} (z) K_{i\frac{\Omega}{a} - \frac{1}{2}} (z) = \frac{i \pi^2}{4 \sinh(\frac{\pi}{2a}(\Omega' - \Omega)) \cosh(\frac{\pi}{2a}(\Omega' + \Omega))}~,
	\label{A13}
\end{equation}
using the integral relation 6.576(4) for modified Bessel functions from \cite{Gradshteyn2007}:
\begin{eqnarray}
&\int_{0}^{\infty} dz \, z^p K_v (z) K_{v'} (z) =\frac{1}{2^{2-p} \Gamma(1+p)} \times\nonumber \\
&	 \left[ \Gamma (\frac{1 + p + (v + v')}{2}) \Gamma(\frac{1+p - (v+v')}{2}) \Gamma (\frac{1 + p + (v-v')}{2}) \Gamma( \frac{1+p - (v-v')}{2}) \right]~,
\end{eqnarray}
along with the following Gamma function relations \cite{Gradshteyn2007}:
\begin{equation}
	\begin{aligned}
	\Gamma(z^*) &= \Gamma^*(z)~, \\
	z\Gamma(z) &= \Gamma(z+1)~, \\
	|\Gamma(iz)|^2 &= \frac{\pi}{z \sinh(\pi z)}~, \\
	|\Gamma(\frac{1}{2} + iz)|^2 &= \frac{\pi}{\cosh(\pi z)}~, \\
	|\Gamma(1 + iz)|^2 &= \frac{\pi z}{\sinh(\pi z)}~.
	\end{aligned}
\end{equation}
The integral (\ref{A13}) is one of two terms that arise from the normalisation condition (\ref{A12}). The second term is obtained by taking $\Omega \rightarrow \Omega'$. The total contribution is identically 0, except at the pole $\Omega = \Omega'$. 
We can evaluate the pole by appealing to the Sokhotski-Plemelj formula. In a distributional sense it can be shown that:
\begin{equation}
	\lim_{\varepsilon \rightarrow 0} \frac{1}{\sinh(\frac{\pi}{2a}(\Omega' - \Omega)) + i\varepsilon} = -i2 a \delta(\Omega' - \Omega) + \frac{\rm P.V.}{\sinh(\frac{\pi}{2a}(\Omega'-\Omega))},
	\label{A16}
\end{equation}
where ${\rm P.V.}$ stands for the Cauchy Principal Value. By combining (\ref{A11}) - (\ref{A13}) with (\ref{A16}) we find the normalisation constant: 
\begin{equation}
	N_\Omega = \left( \frac{m \cosh(\frac{\Omega}{a}\pi)}{\pi^2 a} \right)^{\frac{1}{2}}~,
\end{equation}
which gives us the mode solutions (\ref{17}).

\subsection{Bogoliubov coefficients}
Here we present the explicit calculations only for the $\alpha_\Omega$ coefficient, as the $\beta_\Omega$ coefficient is obtained trivially by changing the sign of $\Omega$ as in (\ref{23}). 
Utilizing the normalisation condition on the mode functions we can rewrite (\ref{21}) as
\begin{equation}
	\alpha_{\Omega,k} = \int_{0}^{\infty} d\chi \; {\psi_\Omega^{\dagger}}^{(\pm)} \Lambda \psi_{\omega,k}^{(\pm)}.
\end{equation}
Written out explicitly, this is:
\begin{equation}
	\int_{0}^{\infty} d\chi \left( \frac{m \cosh(\frac{\Omega\pi}{a})}{\pi a} \right)^{\frac{1}{2}} e^{i\Omega \tau} \begin{pmatrix}  K_{-i\frac{\Omega}a + \frac{1}{2}} \\ -i K_{-i\frac{\Omega}{a} - \frac{1}{2}} \end{pmatrix}^{T} e^{-\frac{1}{2} a \tau} \frac{1}{(2\pi)^{\frac{1}{2}} \sqrt{2 \omega_p}} e^{-i \omega_p \chi \sinh(a\tau) + ip \chi \cosh(a\tau)} \begin{pmatrix} \sqrt{\omega_p -p} \\ \sqrt{\omega_p + p} \end{pmatrix}.
\end{equation}
We then use the change of variables:
\begin{equation}
		\omega = m \cosh(\theta_p)~,~~ k = m \sinh(\theta_p)~, 
\end{equation}
to obtain:
\begin{equation}
	\alpha_{\Omega,\theta_p} = \int_{0}^{\infty} \frac{d \chi}{2\pi} \sqrt{ \frac{m \cosh(\frac{\Omega\pi}{a})}{\pi a \omega_p}} e^{i\Omega \tau} e^{-im \chi \sinh( a\tau - \theta_p)} \sqrt{m} \left[ K_{-i\frac{\Omega}{a} + \frac{1}{2}}(m\chi) - i K_{-i\frac{\Omega}{a} + \frac{1}{2}}(m\chi)e^{-\frac{1}{2}(a\tau - \theta_p)}  \right]~.
\end{equation}
This can be solved using the integral 6.611(3) \cite{Gradshteyn2007}:
\begin{equation}
	\int_{0}^{\infty} e^{-\alpha z} K_v (\beta z) = \frac{\pi \text{cosec}(v \pi)}{2 \sqrt{\alpha^2 - \beta^2}} \left[ \left( \frac{\alpha + \sqrt{\alpha^2 - \beta^2}}{\beta} \right)^v - \left( \frac{\alpha + \sqrt{\alpha^2 - \beta^2}}{\beta} \right)^{-v}\right]~.
\end{equation}
After some algebra, we obtain: 
\begin{equation}
	\alpha_{\Omega, \theta_p} = \frac{1-i}{2} \frac{e^{\frac{i \Omega \theta_p}{a}} e^{\frac{\pi \Omega}{2a}}}{\sqrt{2a\pi m \cosh(\theta_p) \cosh(\frac{\Omega}{a} \pi) }}~.
\end{equation}

\subsection{Rindler current and charge density calculation}

We are interested in calculating the expectation values $\langle1_p | \hat j^\mu  | 1_p\rangle$, where $\hat j^{\mu}$ is given in (\ref{19},\ref{20}). Explicitly, we have: 
\begin{equation}
	\begin{aligned}
		\left< 1_p | :\hat{{\psi}}^\dagger \hat O \, \hat{\psi} :| 1_p \right> =&  \int_{-\infty}^{\infty} \int_{-\infty}^{\infty} d\Omega \, d\Omega' \left({\psi^{(+)}_\Omega}\right)^\dagger \hat O \, \psi^{(+)}_{\Omega'} \; \left[	\alpha_{\Omega,\theta_p}^* \alpha_{\Omega',\theta_p} \, + \int_{-\infty}^{\infty} dk \, \alpha_{-\Omega,\theta_k}^* \alpha_{-\Omega',\theta_k} \delta(0) \right] \\
		&-2 \int_{0}^{\infty} \int_{0}^{\infty} d\Omega \, d\Omega' \int_{-\infty}^{\infty} dk \; \left({\psi^{(-)}_\Omega}\right)^\dagger \hat O \, \psi^{(-)}_{\Omega'} \; \alpha_{-\Omega,\theta_k}^* \alpha_{-\Omega',\theta_k} \delta(0) ~,	\label{A31}
\end{aligned}
\end{equation}
where $\hat O$ is $\mathbbm{1}$ for the charge density and $\gamma^0 \gamma^1 a \chi$ for the current density. 
The third term arises from the normal ordering of the Rindler operators and exactly cancels with the divergent second term (this can be shown by using the relationship between $\alpha_\Omega$ and $\beta_\Omega$, along with the standard normalisation of Bogoliubov coefficients). The first term is the only contribution to this expectation value. 
For the current density, this integral is
\begin{equation}
	\int_{-\infty}^{\infty}	\int_{-\infty}^{\infty} d\Omega \, d\Omega' \, \alpha^*_{\Omega,\theta_p} \,\alpha_{\Omega,\theta_p}\; \left(\psi^{(+)}_\Omega \right)^\dagger \gamma^0 \gamma^1 a \chi\, \psi^{(+)}_{\Omega'}. \label{A32}
\end{equation} 
Writing this out explicitly we have:
\begin{equation}
	\int_{-\infty}^{\infty} \int_{-\infty}^{\infty} d\Omega \, d\Omega' \, \frac{e^{i\tau(\Omega - \Omega')}\chi e^{-i \frac{\theta_p}{a}(\Omega - \Omega')} e^{\frac{\pi}{2a}(\Omega' + \Omega)}}{4 \pi^3 a \cosh(\theta_p)} \left[ K_{++} K_{+-}^{'} - K_{+-}K_{++}^{'} \right],
\end{equation}
where we have used the shorthand notation $K_{\pm \mp} = K_{\pm i\frac{\Omega}{a} \mp \frac{1}{2}}(m\chi),~ K_{\pm \pm} = K_{\pm i\frac{\Omega}{a} \pm \frac{1}{2}}(m\chi)$ and primes indicate that the order of the Bessel function is a function of $\Omega'$.
We now use the relation 6.664(6) from \cite{Gradshteyn2007}:
\begin{equation}
	K_\alpha(x) K_\beta(x) = 2 \int_{0}^{\infty} dy \cosh(y (\alpha + \beta)) K_{\alpha-\beta} (2x \cosh(y)),
\end{equation}
as well as the Bessel recurrence relation 8.486(10) \cite{Gradshteyn2007}:
\begin{equation}
	K_{\alpha+1} (x) - K_{\alpha-1} (x) = \frac{2\alpha}{x} K_{n}(x)	,
\end{equation}
to show that:
\begin{equation}
	K_{++}K_{+-}^{'} - K_{+-}K_{++}^{'} = 2 \int_{0}^{\infty} dy \frac{i(\Omega - \Omega')}{am \chi \cosh(y)} K_{\frac{i}{a}(\Omega-\Omega')} (2m\chi \cosh(y)) \cosh\left(i \frac{y}{a} (\Omega+ \Omega')\right)~.
\end{equation}
After applying the change of coordinates
\begin{equation}
	\begin{aligned}
		\Omega &= \frac{1}{2} \left( \Omega_+ + \Omega_- \right), \\
		\Omega' &= \frac{1}{2} \left( \Omega_+ - \Omega_- \right), 
	\end{aligned}
\end{equation}
the total integral becomes:
\begin{equation}
	\int_{-\infty}^{\infty} \int_{-\infty}^{\infty} \int_{0}^{\infty} d\Omega_- \, d\Omega_+ \, dy\, \frac{i\Omega_-}{4 \pi^3 a^2 m \cosh(\theta_p)\cosh(y)} K_{\frac{i}{a}\Omega_-} (2m \chi \cosh(y)) \cosh\left(i \frac{y}{a} \Omega_+\right)  e^{i\tau \Omega_-} e^{-i \frac{\theta_p}{a}\Omega_-} e^{\frac{\pi}{2a}\Omega_+}.
\end{equation}
The $\Omega_+$ integral gives two delta functions, specifically:
\begin{equation}
	\pi \delta\left(\frac{y}{a} + \frac{i\pi}{2a}\right) + \pi \delta\left(\frac{-y}{a} + \frac{i\pi}{2a}\right)~.
\end{equation}
The $\Omega_-$ integral is of the form:
\begin{equation}
	\int_{-\infty}^{\infty} \frac{i}{a} \Omega_- K_{\frac{i}{a}\Omega_-} (2m\chi \cosh(y)) i \sin\left(\frac{\Omega_-}{a}(\theta_p + a \tau)\right)~, 
\end{equation}
which can be solved using 6.795(3) \cite{Gradshteyn2007}:
\begin{equation}
	\int_{-\infty}^{\infty} x \sin(ax) K_{ix} (b) dx = \pi b \sinh(a) e^{-b\cosh(a)}.
\end{equation}
After solving both the $\Omega_-$ and the $\Omega_+$ integrals we obtain: 
\begin{equation}
	\int_{0}^{\infty}  dy \frac{a\left[ \delta(y + i\frac{\pi}{2}) + \delta(y - i\frac{\pi}{2}) \right]}{2 \pi^2 \cosh(\theta_p)} \pi  \chi \sinh(\theta_p - a \tau) e^{-2m\chi \cosh(y)\cosh(\theta_p+a\tau)}.
\end{equation}
In order to take the integral over $y$, we deform the integral to the line $z = \pm ia + y$ to integrate the complex arguments of the delta function.
The final result is
\begin{equation}
	\frac{a \chi \sinh(\theta_p - a\tau)}{\pi \cosh(\theta_p)}	.
\end{equation}
Following an almost identical procedure, the charge density (\ref{28}) can be found. 

\subsection{Mode Solutions for the Cosmological Spacetime}

The general solution to the equation of motion, given in \eqref{eq:cosmoeom} can be written in the form

\begin{equation*}
	\psi^+_p = a(\eta)^{-\frac{1}{2}} \begin{pmatrix} \alpha \, U_1(\eta) + \beta \, U_1(\eta)[\omega_{in} \rightarrow -\omega_{in}] \\ \alpha \, U_2(\eta) + \beta \, U_2(\eta) [\omega_{in} \rightarrow -\omega_{in}] \end{pmatrix} e^{ipx},
\end{equation*}
with 
\begin{equation}
	\begin{aligned}
		U_1 = & {}_2F_1 \left[ \frac{i}{\varepsilon}( \omega_- - m \chi) , 1 + \frac{i}{\varepsilon}( \omega_- + m\chi) ; (1- \frac{i}{\varepsilon} \omega_{in}) ; (\frac{1}{2} \tanh (\varepsilon \eta) + \frac{1}{2})\right]  \\\times  & (e^{\varepsilon \eta} + e^{-\varepsilon \eta }) 2^{-\frac{i \omega_{in}}{\varepsilon}} e^{\frac{i \pi}{2}(1 + \frac{i \omega_{out}}{\varepsilon})} e^{- i \omega_+ \eta} e^{-(1 + \frac{i \omega_-}{\varepsilon}) \ln (2 \cosh (\varepsilon \eta))} 
\end{aligned}
\end{equation}
and
\begin{equation}
	\begin{aligned}
		U_2 = &\frac{1}{p}  (e^{\varepsilon \eta} + e^{-\varepsilon \eta }) 2^{-\frac{i \omega_{in}}{\varepsilon}} e^{\frac{i \pi}{2}(1 + \frac{i \omega_{out}}{\varepsilon})} e^{- i \omega_+ \eta} e^{-(1 + \frac{i \omega_-}{\varepsilon}) \ln (2 \cosh (\varepsilon \eta))} \\
		& \times \;\Big[ F_1 \left[ - ma_0 + \tanh(\varepsilon \eta) (i \varepsilon - m \chi) + \frac{i}{2}\left[ (i \omega_{in} - \varepsilon)(\tanh(\varepsilon \eta) -1) - (i \omega_{out} + \varepsilon)(\tanh(\varepsilon \eta) + 1) \right] \right] \\
		& - \frac{F_2}{i \omega_{in} - \varepsilon} \left[4 i\left[ \omega_-^2 - m^2 \chi^2 \right] + 4 \varepsilon(\omega_- - m\chi) \right] \left( \tanh(\varepsilon \eta) +1 \right) \left( \tanh(\varepsilon \eta) -1  \right) \Big].
	\end{aligned}
\end{equation}
In order to find the in/out states, we match this general solution to the plane wave solutions in the limits $\eta \rightarrow -\infty$ and $\eta \rightarrow \infty$ respectively. The plane wave mode solutions to for a spinor field in the 2D Dirac representation are:
\begin{equation}
	\begin{aligned}
		\psi^+_{in/out} = \frac{a(\eta)^{-\frac{1}{2}}}{\sqrt{4 \omega_{in/out} \pi}} \begin{pmatrix} \sqrt{\omega_{in/out} + m a_{in/out}} \\ \frac{|p|}{p} \sqrt{\omega_{in/out} - m a_{in/out}} \end{pmatrix} e^{-i \omega_{in/out} \eta + ipx}, \\
		\psi^-_{in/out} = \frac{a(\eta)^{-\frac{1}{2}}}{\sqrt{4 \omega_{in/out} \pi}} \begin{pmatrix} \frac{|p|}{p} \sqrt{\omega_{in/out} - m a_{in/out}} \\  \sqrt{\omega_{in/out} + m a_{in/out}} \end{pmatrix} e^{i \omega_{in/out} \eta - ipx} .
	\end{aligned}
\end{equation}
The in and out states given in \eqref{eq:instate} and \eqref{eq:outstate} approach these solutions in their respective limits. 

\subsection{Cosmological Bogoliubov Coefficients}
The Bogoliubov coefficients can be obtained by determining how the in state evolves as $\eta \rightarrow \infty$. To do this we use the following hypergeometric function relation\cite{Gradshteyn2007}: 
\begin{equation}
	\begin{aligned}
		F(\alpha, \beta; \gamma; z) =& \frac{\Gamma(\gamma) \Gamma(\gamma - \alpha -\beta)}{\Gamma(\gamma - \alpha) \Gamma(\gamma -\beta)} F(\alpha, \beta; \alpha + \beta - \gamma + 1; 1-z)  \\
		&+ (1-z)^{\gamma-\alpha-\beta} \frac{\Gamma(\gamma) \Gamma(\alpha + \beta-\gamma)}{\Gamma(\alpha)\Gamma(\beta)} F(\gamma-\alpha,\gamma-\beta;\gamma-\alpha-\beta+1; 1-z)	,
	\end{aligned}
\end{equation}
which, when applied to the in state hypergeometric function (in the first component), gives:
\begin{equation}
	\begin{aligned}
		&{}_2F_1 \left[ \frac{i}{\varepsilon} (\omega_- - m \chi) , 1 + \frac{i}{\varepsilon} (\omega_- + m\chi) ; \left(1- \frac{i}{\varepsilon} \omega_{in}\right) ; \left(\frac{1}{2} \tanh (\varepsilon \eta) + \frac{1}{2}\right)\right]  \\
		=& \frac{\Gamma(1 - \frac{i \omega_{in}}{\varepsilon}) \Gamma(-\frac{i \omega_{out}}{\varepsilon})}{\Gamma(1 - \frac{i}{\varepsilon}(\omega_+ - m\chi) \Gamma(-\frac{i}{\varepsilon}(\omega_+ + m\chi))} {}_2F_1\left[   \frac{i}{\varepsilon} (\omega_- - m \chi) , 1 + \frac{i}{\varepsilon} (\omega_- + m\chi) ; 1 + \frac{i \omega_{out}}{\varepsilon}; \left(\frac{1}{2} - \frac{1}{2} \tanh (\varepsilon \eta)\right)\right] \\
		&+ e^{i\omega_{out} \eta} e^{\frac{i\omega_{out}}{\varepsilon} \ln 2 \cosh \varepsilon \eta} \frac{\Gamma(1 - \frac{i \omega_{in}}{\varepsilon}) \Gamma(\frac{i \omega_{out}}{\varepsilon})}{\Gamma(\frac{i}{\varepsilon}(\omega_- - m\chi) \Gamma(1 +\frac{i}{\varepsilon}(\omega_- + m\chi))}\\
		&\times {}_2F_1\left[1 -   \frac{i}{\varepsilon} (\omega_+ - m \chi) , -\frac{i}{\varepsilon} (\omega_+ + m\chi) ; 1 - \frac{i \omega_{out}}{\varepsilon}; \left(\frac{1}{2} - \frac{1}{2}\tanh (\varepsilon \eta)\right)\right] ~,
	\end{aligned}
\end{equation}
which allows us to write
\begin{equation}
	\psi_{in}^+ = \alpha_p \psi_{p, out}^+ + \beta_p \psi_{-p, out}^-
\end{equation}
with $\alpha_p$ and $\beta_p$ defined as in \eqref{eq:cosmoalpha} and \eqref{eq:cosmobeta}.

\end{document}